\documentclass{epl}

\title{The Fractal Properties of Internet}
\author{G. Caldarelli\inst{1} \and R. Marchetti\inst{1} \and L. Pietronero\inst{1}}
\institute{
  \inst{1} INFM Unit\`a ROMA1 and Dip. di Fisica - 
Universit\`a di Roma "La Sapienza", P.le A. Moro 2, 00185 Roma, Italy
}

\begin{document}

\maketitle

\begin{abstract}
In this paper we show that the Internet web, from a user's perspective, 
manifests robust scaling properties of the type $P(n)\propto n^{-\tau}$ 
where n is the size of the basin connected to a given point, $P$ represents 
the density of probability of finding n points downhill and $\tau=1.9 \pm 0.1$ s a 
characteristic universal exponent. This scale-free structure is a result of 
the spontaneous growth of the web, but is not necessarily the optimal one for 
efficient transport. We introduce an appropriate figure of merit and suggest 
that a planning of few big links, acting as information highways, may 
noticeably increase the efficiency of the net without affecting its robustness.
\end{abstract}

Networks are present in many aspects of everyday life, from the watershed where 
the rivers water is collected, to the veins and lymphatic channels that 
distribute blood and nutrition in animals and plants \cite{1,2}, to the telephone or 
electricity or internet webs that transport in our houses the services we need.
In all these cases, the network properties should be such to optimise some cost 
function, as for example the number of points connected with respect to the  
length of the web. In this paper we analyze the structure of the Internet web. 
The connections between users and providers are studied and modeled as branches 
of a world spanning tree. These results have scientific and technological 
implications that are briefly described. In addition we propose a model based 
on a stochastic Cayley tree which accounts for both qualitative and quantitative 
properties and can be used as a prototype model to explore and optimise the 
characteristics of the system.
The model is inspired to the theory of river networks \cite{3}, which can 
provide an explanation of the fractal properties of the net with respect to the 
optimization of some thermodynamic potential. This question is not only of a 
scientific relevance, but it also addresses a very important technological 
question. Namely, which cost function has to be minimised in order to improve 
the net properties both to plan future wiring of developing countries and to 
improve the quality of the net connection for countries already connected. 
For network formation, Nature often chooses fractal structures. 
Fractal objects introduced by Benoit Mandelbrot\cite{4} are characterised by 
the property of having similar properties at all length scales. In this respect 
they show the same complexity at different scales without a characteristic 
scale or size for their structures. These properties are defined between a 
lower and an upper scales which, for the present case, are the size of a 
single node and the total world network. It is exactly this scaling property 
that allows animals to survive with a quantity of blood much smaller than the 
solid volume occupied by their body. The fractal structure of veins distributes 
the blood so efficiently that every cell is reached in a reasonably short path 
with the minimum possible structure.
This paper addresses the issue of the characterization and the design of a 
rational and optimal web for Internet by using the examples present in Nature 
for similar structures. The statistical study of the Internet has already 
started to attract the interest of the scientific community \cite{5,6,7,fal}. 
Differently from these studies we focus here on the physical layer of the net, 
rather than on the structure of symbolic links in the web. A striking evidence 
of the network structure of this physical layer has been provided by the wonderful 
maps realised in the ``Internet Mapping Project'' by William R. Cheswick, Hal Burch and 
Steve Branigan at the Bell Laboratories
\footnote{A description of the project, together with maps done by Cheswick B. \and Burch H.,
  is available at http://www.cs.bell-labs.com/who/ches/map/index.html}.

\begin{figure}
\onefigure{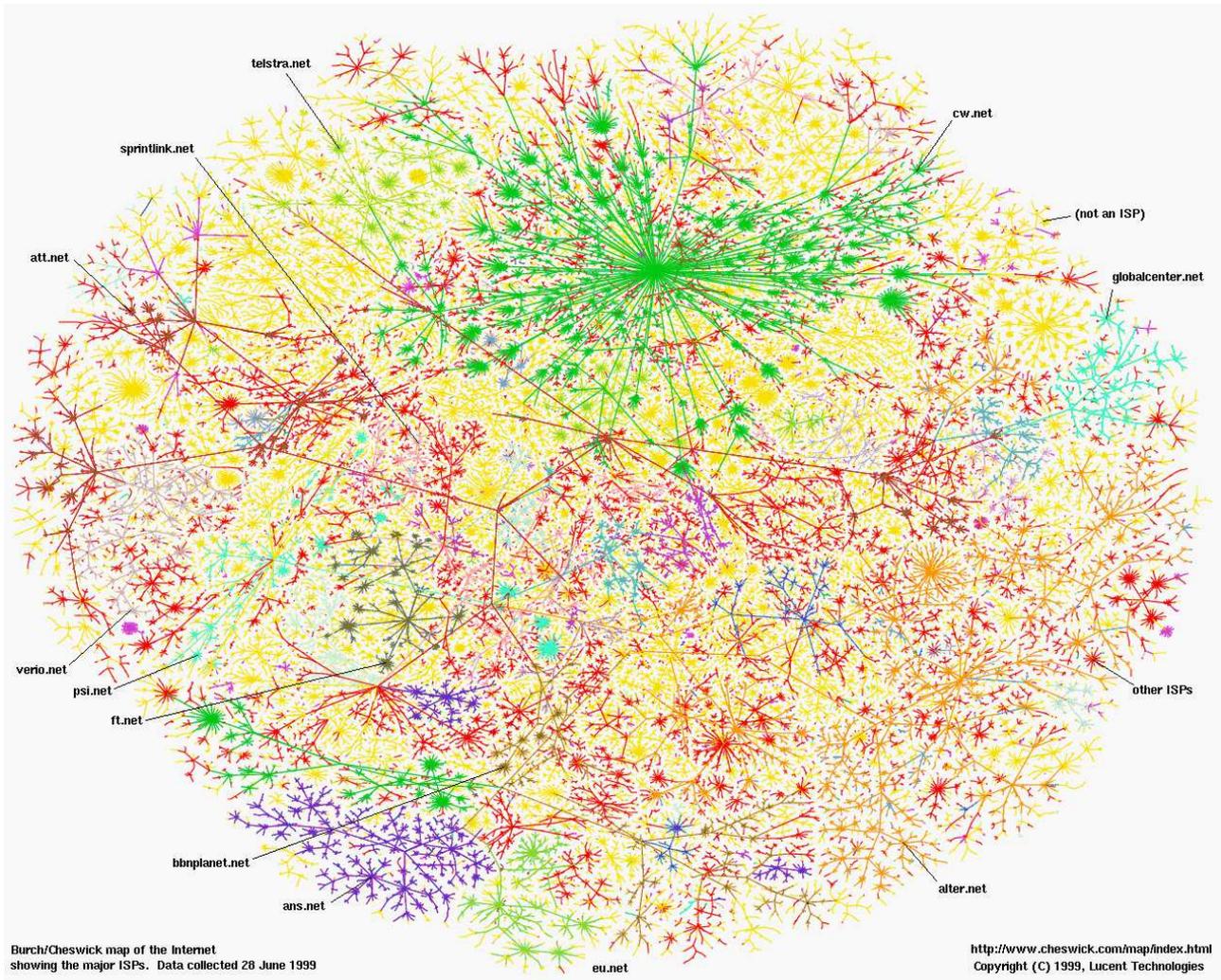}
\caption{Picture of the net realised in Bell Laboratories, Courtesy of Lucent Technologies. 
This map represents the net as seen by a single user.}
\label{f1}
\end{figure}

\begin{figure}
\onefigure{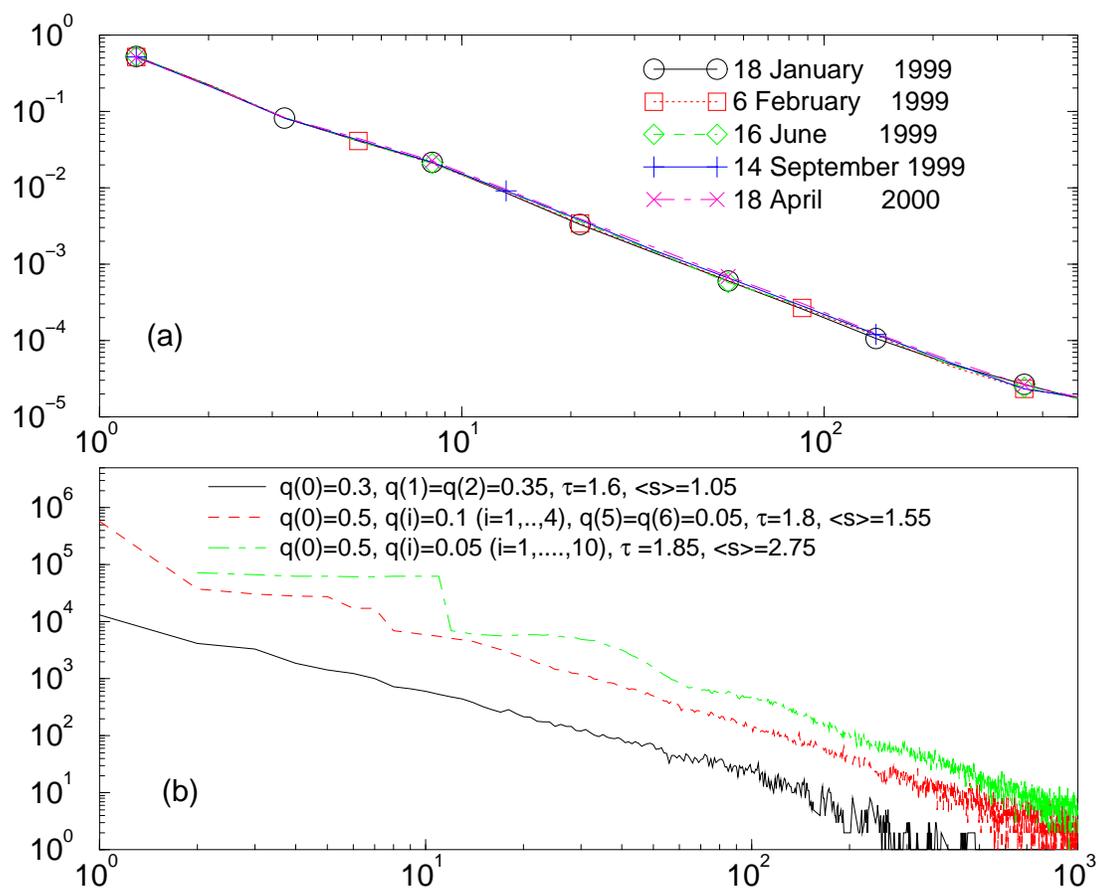}
\caption{(a) Plot of the density function $P(n)$ of the whole network of $1249747$ nodes. 
This function gives the probability density of having n descendants (basin of attraction). 
(b) Plot of the Density Function $P(n)$ for different Cayley trees, characterised by different 
distributions of sons per site. 
These plots correspond to averages over $500$ realizations with a maximum number of generations 
equal to $25$.}
\label{f2}
\end{figure}

One of these maps is reproduced 
in fig.~\ref{f1}, showing the great complexity of this sample of paths on the physical network of 
Internet cables. These maps are drawn from  a set of data obtained through a computer 
instruction that allows to trace the route from one terminal to any allowed address 
in the Internet domain. This command records all the nodes through which the target 
is reached from the starting point. These paths actually change over time for the 
following reasons: firstly the routes reconfigure since the path is variable according 
to the traffic at the moment or more generally according to the availability of the 
connection. Secondly the whole structure is physically evolving due to new connections 
that take place. Furthermore, since the collection of data takes some time, it is possible 
that sometimes some loops are formed in the net, since in principle the track through 
the same point can be different at different times. We checked that this is an unlikely 
event and when this happens, most of the times the different branches of a loop share 
similar statistical properties. In this perspective, these sets of data are then suitable 
to be studied with the same framework introduced for river networks. In particular one 
can consider the terminal located in the Bell Laboratories as the outlet of a river 
basin, and the path connecting this point to all the possible net addresses can be 
considered to form 
the structure of this basin. It is then interesting to measure this basin by using the 
density function $P(n)$ expressing the probability that a point in the structure connects 
$n$ other points uphill. Such a quantity, also known as the drainage area, represents the 
number of points that lie uphill a certain point in the net. As a signature of the 
intrinsic fractal properties of webs this density function for self-similar objects is a 
power law, that is $P(n)\propto n^{-\tau}$. The value of this exponent $\tau$ 
allows one to address and to distinguish between different physical features. 
As regards the river basins, for example, 
striking similarities can be noted amongst the river structures in all the world. Namely, 
the interplay between soil erosion and drainage network conduces the system towards a state 
where the total gravitational energy that is dissipated is minimal\cite{9,10}. 
This universality 
accounts for the fact that, regardless the landscape peculiarities, the optimal solution to 
the drainage problem must be the same everywhere1. These optimal structures are characterised 
by an exponent $\tau=1.45$, different from that corresponding to a set of random spanning 
trees (for which $\tau=11/8=1.375$ \cite{11}) and from the extremal ones 
(for which $\tau=3/2=1.5$ \cite{12}). 
For the internet data we perform the same analysis of the statistical measure as shown 
in fig.~\ref{f2}(a), which characterises the network's properties as seen from the perspective of 
a user. The results show a clear power law, extending over more than two decades, whose 
exponent $\tau$ is equal to $1.9 \pm 0.1$,
 for different measurements realised from the  18-01-1999 to the 18-04-2000. This value of 
the exponent represents a first important difference with the theory of river networks where 
a maximum value of $\tau=1.5$ is supposed \cite{13}. 
It should be possible to understand such a difference by the different metric 
nature of the space considered for points in watershed and sites in space of the addresses. 
Since new nodes appear every day and connect to existing providers, the whole dimension of 
the basin is changing with time in a hierarchical rather than spanning way. This means that 
space can become oversatured with respect to the river network case. From this property and 
from the analysis that has been carried out on the hierarchical structure of the web 
pages\cite{5,7} it appears natural to describe 
this hardware structure in terms of a hierarchical pattern of the kind of the Cayley tree 
with some randomness in the process of elementary bifurcation. This hierarchy of levels is 
believed to play an important role for the network properties and it leads to specific features 
which cannot be reproduced by other models. For example Barabasi and
Albert\cite{6} have shown 
that models with random connections but without a hierarchical structure are not fractal in the 
sense that the probability of finding highly connected site is decreasing exponentially with the 
number of connections, thereby determining a characteristic scale for the network. We test our 
hypothesis on the hierarchical nature of connections, by defining a tree-like structure departing 
from the outlet of the basin that can be described as a Random Cayley Tree. 
A Cayley tree branches at each generation in k different sites. It is easy to check that deterministic 
Cayley trees have a statistics such that $P(n)\propto n^{-1}$. 
Random behavior can be introduced by assigning a 
probability $q(s)$ that a site will have $s$ sons. Provided that the average number of descendants 
\begin{equation}
\langle s \rangle=\sum_{s=1}^{s_{max}} sq(s) 
\end{equation}
is greater or equal to one there is a finite probability that the tree will last 
forever. The result is dependent on the mean number of sons generated in the evolution, 
starting from a value of $\tau=3/2$ \cite{de} known to be exact when $\langle s \rangle =1$. 
One can measure $\tau=1.6 \pm 0.1$ for a $q(0)=0.3,q(1)=q(2)=0.35$ 
but if one increases the value of $\langle s\rangle$ 
until for example $\langle s \rangle=2.75, q(0)=0, q(1)=....=q(10)=0.05$ 
one can measure the value of $\tau=1.85 \pm 0.1$ in good agreement with the 
real data (fig.~\ref{f2}(b)). Since it is possible to show that in case of increasing population for stochastic 
Cayley trees (i.e. $\langle s \rangle >1$) the expected exponent is $\tau=2$\cite{de}, 
we justify the above variation 
as finite size effects in our computer simulations. As a matter of fact, due to these finite size 
effects the convergence to this exponent is usually rather slow and this transient behavior may 
appear as an effective exponent $\tau$ lying between the two exact values $3/2$ and $2$. 
A striking clue of the hierarchical 
structure of the network can also be witnessed by the value of the exponent characterizing the density 
function $P(n)$ of sites uphill a node in the network. Indeed in the case of river networks for a lattice 
spanning network with no organization one has to expect $\tau=1.375$, whilst for a stochastic Cayley tree 
(if the system does not die out) the exponent has to be significatively larger. In addition, from the 
perspective of our model, we predict that, as the number of connections continues to grow, one should 
observe an increase of the exponent $\tau$ towards the asymptotic value $2$.
At this point we are in condition to evaluate in which way it could be possible to optimise the 
network in order (for example) to reduce the mean number of jumps any client has to realise in order 
to be connected with all the other points in the net. A possible solution to this requirement could 
be represented by a particular type of random graph model known as the "small-world"
model\cite{15} 
introduced in order to describe all the social situations where one can get from a member of a 
network to any other member via a small number of intermediate acquaintances. A small world system 
in the simplest version consists of a $1$-dimensional system of $L$ sites with periodic boundary 
conditions (a ring). A site is connected to $k$ different neighbours chosen from the nearest one. 
Furthermore, m shortcuts are present from different randomly chosen sites. If for example  $L=50$, 
$k=3$, and $m=9$ one can evaluate numerically that  the average vertex-vertex distance is of the order 
of $4\div 6$ steps.  One can see the small-world model in terms of transportation. The small local bonds 
represents motion by trains or car, while the big jumps realised by  shortcuts  correspond to 
airplane flights. In this respect there are characteristic lengthscales and the optimal system is 
not scale invariant. A figure of merit can be the average number of steps to connect two random 
points which is typically rather small in the small-world model.  To optimise in such a way the 
Internet network would be highly desiderable. Unfortunately, due to the fractal properties we 
measured on the web, the mean vertex-vertex distance is much larger (around $15$) and furthermore show 
no characteristic cutoff. Namely, the optimization in the distance between vertex is linked to 
the form of the distribution $P(n)$ that for such system is now peaked around a mean value. In this 
case then the $P(n)$ shows no scale-free behaviour (On the other hand, the small world is known to 
be characterised by a correlation distance $\xi= 1/(\phi kd)^{1/d}$, where 
$\phi=m/L$ is the probability to have 
$m$ shortcuts) . Conversely the fractal scale-free structure of the present web, does not guarantee 
a short number of steps between points, but instead  shows that the probability of a very long 
path is small, but finite.
The natural conclusion is therefore that it should be possible to improve the efficiency of 
the net by planning a certain number of big links which should play the role of the shortcuts
 in the small-world model. These information highways superimposed on the network structure 
should play the role of the plane transport without affecting the local structure. 
We believe that our analysis, providing a quantitative measure of the physical Internet network 
could be the starting point of this study. We conclude that, from the perspective of a single 
user (the node from which the route has been traced to all the other nodes), one observes 
essentially a stochastic Cayley tree whose properties are robust with respect of the different 
possible routes used in different tests. This denotes an underlying hierarchical structure that 
links providers and users (this is topologically different from a random spanning tree, because 
now the space is filled in a "thicker" way). We are presently studying this problem in order to 
check  how the present network could be improved by creating suitable shortcuts that can reduce 
in a significant way the cost of connections between different users. 
\acknowledgments
We acknowledge fruitful discussions with  E. Bonabeau, P. De Los Rios, A. Flammini, G.Parisi and F. Rossi

Correspondence and requests for materials should be addressed to G.C. \newline
(e-mail: gcalda@pil.phys.uniroma1.it)

\end{document}